\documentstyle[aps,prl,epsfig]{revtex}

\begin{document}




\def\epp{\epsilon^{\prime}}
\def\vep{\varepsilon}
\def\ra{\rightarrow}
\def\ppg{\pi^+\pi^-\gamma}
\def\vp{{\bf p}}\def\ko{K^0}
\def\kb{\bar{K^0}}
\def\al{\alpha}
\def\ab{\bar{\alpha}}
\def\nm{\nu_\mu}
\def\nt{\nu_\tau}
\def\be{\begin{equation}}
\def\ee{\end{equation}}
\def\bea{\begin{eqnarray}}

\vspace{5.0cm}

\title{UP-DOWN ASYMMETRY OF NEUTRAL CURRENT EVENTS AS A DIAGNOSTIC FOR
$\nu_\mu- \nu_{st}$ VERSUS $\nu_\mu-\nu_\tau$ OSCILLATIONS}

\author{JOHN G. LEARNED$^1$,  SANDIP PAKVASA$^1$ 
and J. L.  STONE$^{2}$} 
\address{$^1$Department of Physics \& Astronomy, 
University of Hawaii at Manoa\\
Honolulu, HI  96822  USA \\
and \\
$^{2}$ Department of Physics \\
Boston University \\
Boston, MA 02215}

\maketitle

\begin{center}
{\it Preprint: UH-511-901-98  May 1998}
\end{center}

\vspace{2.0cm}

\begin{center}
Abstract
\end{center}

\abstract{We show that the asymmetry in the neutral current events
(e.g. $\nu N \ra \nu N \pi^0)$ can be used to discriminate between
$\nm - \nt$ and $\nm - \nu_{st}$ mixing as being responsible for the
atmospheric neutrino anomaly.  Specifically, $A_N$ vanishes for $\nm
- \nt$ mixing and is about 2/3$A_\mu$ for $\nm - \nu_{st}$ mixing.}



\newpage

The neutrino oscillation interpretation~\cite{1} of the atmospheric
neutrino seems more and more probable with the new data from
Superkamiokande~\cite{2}.  If this holds up, it is very important to
determine the specific oscillation scenario at work here.  Recently,
it was shown how the up-down asymmetry of the charged current events
with muons and electrons is a strong discriminant for many different
scenarios~\cite{3,4}.  Preliminary results~\cite{5} for the
asymmetry from Superkamiokande favor $\nu_\mu$ oscillating into
$\nu_\tau$ (or $\nu_{sterile})$.  However, this asymmetry does not
distinguish between the possibility of $\nm-\nt$ oscillations and
$\nm-\nu_{st}$ oscillations. The reason for this is that at the
energies in question ($E_\nu < 10 \ GeV$) there are not enough $\nt$
charged current events to be recognised, due to the small
cross-section below about $20 \ GeV$.  Here we extend the asymmetry
to neutral current events (e.g. $\nu N \ra \nu N \pi^0 $)~\cite{6}
and show how this asymmetry can be used to distinguish easily
between the two possibilities.  We assume here that the correlation
between the direction of $\pi^0$ in $\nu N \ra \nu N \pi^0$ and the
initial neutrino direction is strong enough to distinguish up from
down, not a very stringent requirement.

We define the asymmetry as before
\be
A = \frac{D-U}{D+U}
\ee

where D is the number of downward-going events and $U$ is the number
of upward-going events.  A can be defined for both muon charged
current events $A_\mu$ as well for single $\pi^0$ neutral current
events $A_N$.  We assume the detector to be up/down symmetric, and
the data set to be free of significant background.

We now calculate this asymmetry for the  cases of interest:
(i) $\nm - \nt$ oscillations and
(ii) $\nm-\nu_{st}$  oscillations.

\section{$\nm-\nt$ Mixing}

The $\nm$ flux is modified as

\be
N_\mu = N_\mu^0 \ P_{\mu \mu}
\ee

where $P_{\mu \mu} = 1 - \sin^2 2 \theta \sin^2 \left ( \frac{\delta
m^2 L}{4E} \right )$ and $N_\mu^0$ is the flux of $\nm$'s in absence
of oscillations. At very low energies, $P_{\mu\mu} \approx 1 -
\frac{1}{2} \sin^2 2 \theta$ and $A_\mu \ra 0$.  At higher energies,
L/E is negligible for down $\nu_\mu's$ and hence $N_\mu^d=N^0_\mu$.
For upward going $\nu_\mu's$, $N^u_\mu = P_{\mu\mu} N_\mu^0$ and

\be
A_\mu = \frac{1-P_{\mu\mu}}{1+P_{\mu\mu}} \approx
\frac{sin^2 2 \theta}{4 - sin^2 2 \theta}
\ee

which has a maximum of $1/3$ for $\sin^2 2 \theta \approx 1.$

In the $\nm - \nt$ case, since total flux of flavor neutrinos:
$N_{\nu_{e}} + N_{\nu_{\mu}} + N_{\nu_{\tau}}$ does not change, the
neutral current asymmetry $A_N$ is zero.

\section{$\nm-\nu_{st}$ Mixing}

In this case $A_\mu$ is the same as in the $\nm-\nt$ case.  The
asymmetry $A_N$ in neutral current events is given by

\begin{eqnarray}
A_N & = &
\frac{(N^d_{\nu_{e}} + N^d_{\nu_{\mu}}) - (N^u_{\nu_{e}} + N^u_{\nu_{\mu}})}
{(N^d_{\nu_{e}} + N^d_{\nu_{\mu}})      +  (N^u_{\nu_{e}} + N^u_{\nu_{\mu}})}
  \\ \nonumber
& \simeq &
\frac{N^d_{\nu_{\mu}} - N^u_{\nu_{\mu}}}
{(N^d_{\nu_{e}} + N^u_{\nu_{e}}) + (N^d_{\nu_{\mu}} + N^u_{\nu_{\mu}})}\\ 
A_N & \simeq & \frac{A_\mu}{1 + \left \{\frac{2r}{(1+P_{\mu\mu})} \right \}  }
\simeq  \frac{A_\mu}{1 +         \left\{ r (1+A_\mu)      \right \} }
\end{eqnarray}

Here r = $N^0_{\nu_{e}}/N^0_{\nu_{\mu}}$ and for $r \sim 0.45$ to
$0.37 (E_\nu$ upto 5 GeV), and $A_\mu$ near 1/3, $A_N$ is in the
range 0.2 to 0.22.

In Fig. 1 we show a plot of $A_\mu$ and $A_N$ as functions of
$E_\nu$ for $\sin^2 2 \theta \approx 1$ and $\delta m^2 \approx
5.10^{-3} eV^2$. At low energies the small negative asymmetry in the
$\nu_e$ flux due to the earth magnetic field effects makes $A_N$
non-zero in the case of $\nm-\nt$ mixing and dilutes $A_N$ in the
$\nm-\nu_{st}$ mixing. We calculated energy spectra between 0.2 and
5.0 GeV for a detector with an exposure of 22 kiloton-years
(approximately one year of Super-Kamiokande data).  We use the
Bartol flux model~\cite{7}, and a simple quark model for the
cross-sections, and assume a perfect detector.  Detailed
calculations for a particular instrument will of course vary, but
the asymmetry will change little, the general behavior being
insensitive to details.  Matter effects (which are present for
$\nm-\nu_{st}$ mixing but absent for $\nm-\nt$ mixing) and the
angular spread will be discussed elsewhere, but should have no major
effect on the different behavior in the two cases discussed here.  A
different method to distinguish $\nm-\nt$ mixing from $\nm-\nu_{st}$
mixing has been proposed recently~\cite{8}. This technique depends
on the fact that the ratio of the neutral current event rate to the
charged current event rate is quite different in the two cases. The
method we propose here has the advantage that it is independent of
the efficiency for detecting $\pi^0$'s and of the knowledge of
neutral current cross-sections.

\section*{Acknowledgments}

Two of us (S.P. and J.S.) thank Jose Nieves for the hospitality and the
conducive atmosphere at the Tropical Workshop in San Juan where this
work was begun and two others (J.G.L. and S.P.) thank Kazu Mitake for
the conducive atmosphere he provides them in Honolulu.  This work was supported in part
by U.S.Department of Energy.



\begin{figure}[htb]
\begin{center}
\psfig{figure=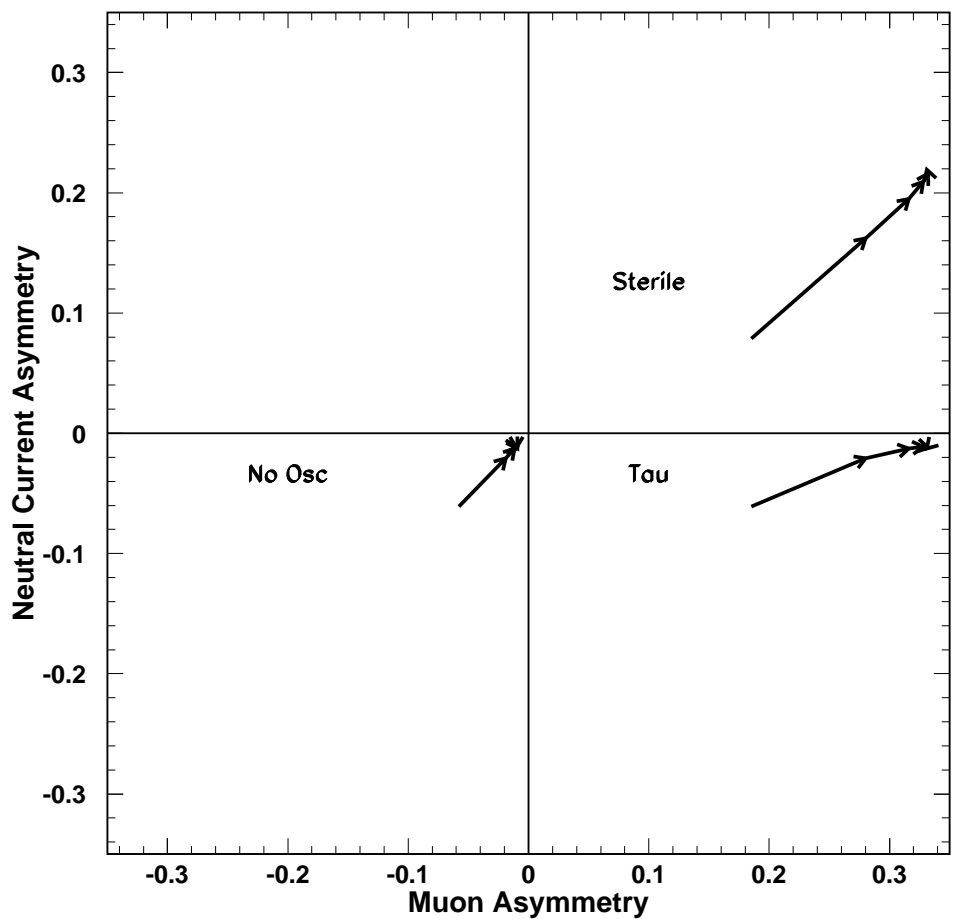,width=14cm}
\caption{ The trajectories of muon asymmetry and neutral current
asymmetry for $\nm - \nt$ mixing, for $\nm - \nu_{st}$, 
mixing, and for no oscillations.}
\label{fig:nc_vs_mu_asym}
\end{center}
\end{figure}
\end{document}